\documentclass[prl,reprint,showpacs,twocolumn,superscriptaddress,floatfix]{revtex4-1}

\usepackage[dvips]{graphicx}
\usepackage{color}
\usepackage{bm}
\usepackage{amsmath}
\usepackage{amsfonts}
\usepackage{bbold}
\usepackage[colorlinks,linkcolor=blue,citecolor=blue]{hyperref}
\usepackage{dsfont}

\newcommand{\average}[1]{\langle #1 \rangle}

\def \ket#1{\mathinner{|{#1}\rangle}}
\def \bra#1{\mathinner{\langle{#1}|}}

\newcommand{\matrixel}[3]{{\mathinner{\langle{#1}| {#2} | {#3}\rangle}} }

\newcommand{\rvec}{{\bf{r}}} 
\newcommand{\kvec}{{\bf{k}}} 
\newcommand{\qvec}{{\bf{q}}} 
\newcommand{\Efvec}{{\bf{E}}_f}

\newcommand{\ketBCS}{\ket{\rm{BCS}} }     
\newcommand{\braBCS}{\bra{\rm{BCS}} }

\begin{document}

\title{Sauter-Schwinger effect in a Bardeen-Cooper-Schrieffer superconductor}

\author{P. Solinas}
\author{A. Amoretti}
\affiliation{Dipartimento di Fisica, Universit\`a di Genova, via Dodecaneso 33, I-16146, Genova, Italy}
\affiliation{INFN - Sezione di Genova, via Dodecaneso 33, I-16146, Genova, Italy}
\author{F. Giazotto}
\affiliation{NEST, Instituto Nanoscienze-CNR and Scuola Normale Superiore, I-56127 Pisa, Italy}


\begin{abstract}
From the sixties a deep and surprising connection has followed the development of superconductivity and quantum field theory.
The Anderson-Higgs mechanism and the similarities between the Dirac and Bogoliubov-de Gennes equations are the most intriguing examples. In this last analogy, the massive Dirac particle is identified with a quasiparticle excitation and the fermion mass energy with the superconducting gap energy.
Here we follow further this parallelism and show that it predicts an outstanding phenomenon: the superconducting Sauter-Schwinger effect (SSSE).
As in the quantum electrodynamics Sauter-Schwinger effect, where an electron-positron couple is created from the vacuum by an intense electric field, we show that an electrostatic field can generate two coherent excitations from the superconducting ground-state condensate.
Differently from the dissipative thermal excitation, these form a new macroscopically  \emph{coherent} and \emph{dissipationless} state.
We discuss how the superconducting state is weakened by the creation of this kind of excitations.
In addition to shed a different light and suggest a method for the experimental verification of the Sauter-Schwinger effect, our results pave the way to the understanding and exploitation of the interaction between superconductors and electric fields.
\end{abstract}

\maketitle

\begin{figure}
    \begin{center}
    \includegraphics[width=\columnwidth]{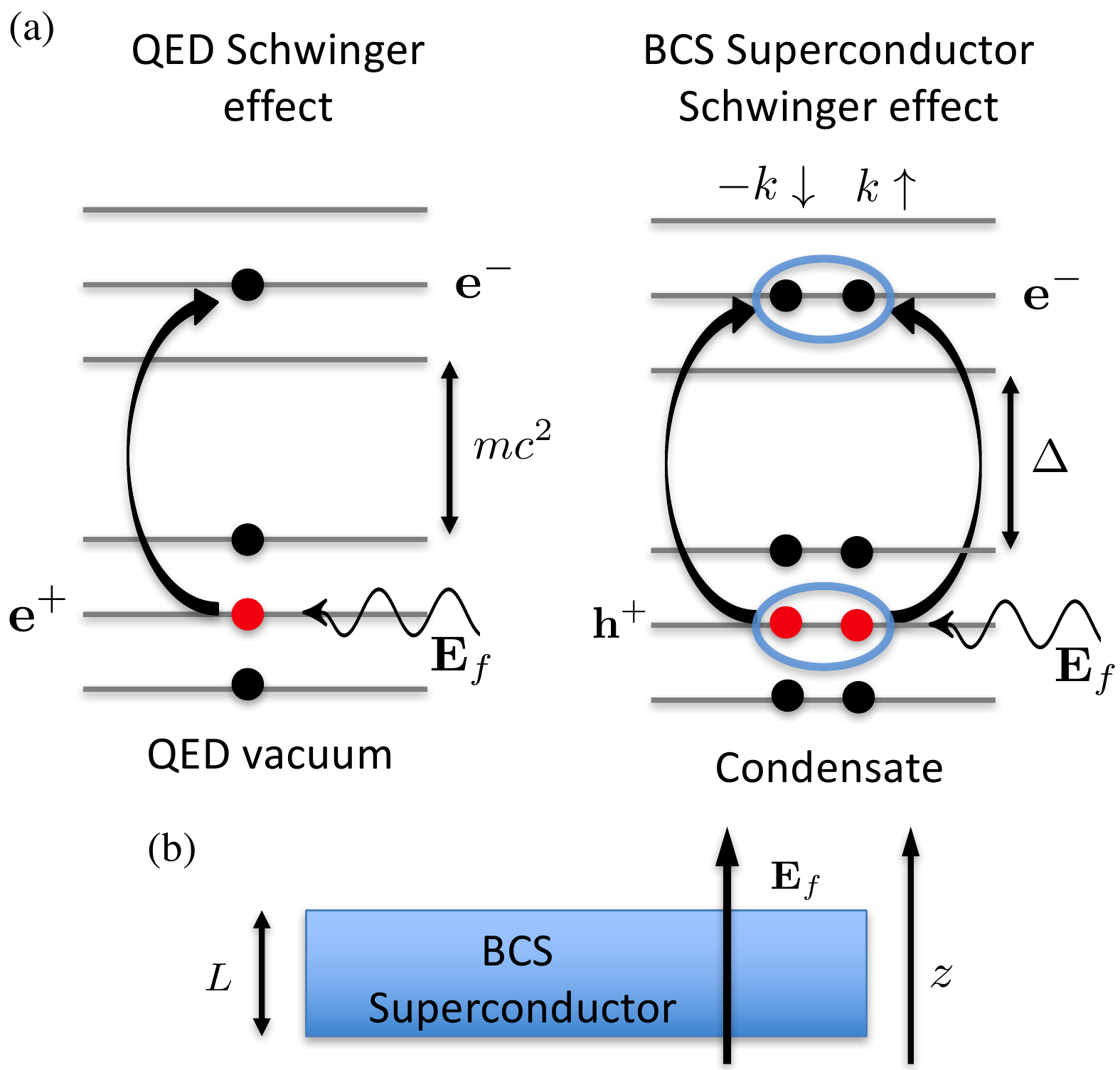}
       \end{center}
    \caption{(a) Analogy between the Sauter-Schwinger effect in QED and in a BCS superconductor. 
(b) Sketch of the experimental setup: a BCS superconducting film of thickness $L$ immersed in an external electric field $\Efvec$.
     }  
    \label{fig:fig1}
\end{figure}

The Sauter-Schwinger effect refers to the creation of an electron and a positron pair from the QED vacuum as a consequence of its instability under the presence of an external electric field [see Fig. \ref{fig:fig1} (a)] \cite{Sauter1931, Heisenberg-Euler}. After the development of quantum electrodynamics (QED), in $1951$ Schwinger gave a complete treatment of the effect \cite{Schwinger1951}. He computed the critical electric field above which the vacuum becomes unstable thereby resulting in the creation of an electron-positron pair. Despite being predicted almost seventy years ago, the Sauter-Schwinger effect has never  been observed so far because of the ultra-high electric fields ($\sim 10^{18}~$V/m) needed. 
In the last twenty years, the Sauter-Schwinger effect has been discussed in different physical frameworks including Mott insulators \cite{Green2005, Oka2003, OkaPRB2012} and QCD  systems \cite{Chernodub2010,Chernodub2011}.

In the sixties,  Nambu and Jona-Lasinio noticed a striking similarity between the Dirac equation and the Bogoliubov-de Gennes equations  describing the elementary excitations in a superconductor \cite{Nambu_Jona-Lasinio1961a,Nambu_Jona-Lasinio1961b,Nambu1960,Aitchison_Hey, deGennes}.
They are formally identical if one identifies the Dirac particle with a quasiparticle excitation in the superconductor and the fermion mass with the superconducting energy gap \cite{Aitchison_Hey}.
In the spirit of the superconductor-QED analogy \cite{Nambu_Jona-Lasinio1961a,Aitchison_Hey}, we would expect to observe a kind of vacuum instability, and a Sauter-Schwinger effect in superconductors exposed to an electric field.
Notably, substituting the electron mass energy ($0.5~$MeV) with the superconducting gap energy ($\sim 100~\mu$eV-$1~$meV for a conventional superconductors), the critical electric field needed to activate the SSSE is drastically reduced. In this perspective, this makes superconductors ideal candidates to realize and measure the Sauter-Schwinger effect.

Despite the appeal and the strength of the superconductor-Dirac particle analogy, in realistic implementations we must keep in mind the important differences between the two physical systems. First, in metallic superconductors (i.e., the ones well-described by the Bardeen, Cooper, and Schrieffer (BCS) theory \cite{deGennes}) screening effects can limit the penetration of the electric field in the interior of the system \cite{VirtanenPhysRevB2019}.
Secondly, a superconductor immersed in an external electric field must be treated like an open quantum system, as the effects of the environment and dissipation processes become relevant \cite{kopnin2001theory,Kopnin2002}.
Regardless of their importance, a detailed analysis of these points would complicate the discussion and drain the attention away from the real focus of the paper. For these reasons, we constraint our discussion to a very specific situation.
We consider a film superconductor thin enough to be completely penetrated by the electric field (or, alternatively, we refer to the effect on the edge of a superconductor), and we analyse the unitary evolution arguing that the environment can affect the dynamics only on longer timescales.
These assumptions allow us to focus on a more precise question: \textit{can a 
static electric field induce a Sauter-Schwinger-like effect in a BCS superconductor by exciting the condensate ground state?}
Below we will show that the answer is indeed affirmative.

Our starting point is the effective Hamiltonian describing a standard BCS superconductor in the presence of an external electric field \cite{deGennes, SM}.
We assume that the electric field $\Efvec = \{ 0, 0,  E_f\}$ is applied to a thin film superconductor along the  $z$ direction [see Fig. \ref{fig:fig1} (b)], and that the superconductor thickness $L$ allows full electric field penetration in the sample.
By decomposing the fermion field in plane waves and choosing the proper gauge, we obtain  the effective Hamiltonian in second quantization formalism \cite{SM, Tsuchiya2018,Tsuji2015, Lancaster-Blundell}
\begin{equation}
 H =
 \sum_\kvec \Big \{
 h_{\kvec_- } (a^\dagger_{\kvec \uparrow} a_{\kvec \uparrow} + a^\dagger_{\kvec \downarrow} a_{\kvec \downarrow}) - \Delta a^\dagger_{\kvec \uparrow} a^\dagger_{-\kvec \downarrow} + h.c. 
 \Big \},
 \label{eq:H_eff_a}
\end{equation}
where $h_{\kvec_\pm } = \frac{1}{2m} \Big[ \hbar^2 k^2_\perp + (\hbar k_z \mp e E_f t)^2 \Big] -\mu$, $k^2_\perp = k^2_x+k^2_y$, $\mu$ is the chemical potential and $m$ is the electron mass.
The order parameter is $\Delta = V \sum_\kvec \average{ a_{\kvec \uparrow } a_{-\kvec \downarrow} }$ and, under this gauge choice, it acquires a superconducting phase $\Delta = |\Delta| e^{i \chi}$ with $\chi = \frac{2 e}{\hbar} E_f t z$ \cite{SM, deGennes,Kopnin2002}.
Notice that, with this gauge choice, the phase factor depends on $z$ so does the Hamiltonian (\ref{eq:H_eff_a}), which is in the standard BCS form thereby allowing a great simplification.  
The price to pay is, however, to deal with a time-dependent problem.

As in the homogeneous case, only $(\kvec, \uparrow)$ and $(-\kvec, \downarrow)$ are coupled. 
Separating the negative $\kvec$ contributions in the kinetic terms of Eq. (\ref{eq:H_eff_a}), we have $h_{-\kvec_- } = h_{\kvec_+ }$
\cite{SM} and, as usual, reversing the momentum is equivalent to change the sign of the particle charge.
Defining $\xi_k(t)= (h_{\kvec_-}+h_{\kvec_+})/2 =(\hbar ^2 k^2+ e^2 E_f^2 t^2)/(2 m)-\mu$
as the kinetic energy (with $k^2 = k^2_\perp + k^2_z$), we can cast Eq. (\ref{eq:H_eff_a}) into a matrix form by means of the Anderson pseudospin representation \cite{Anderson1958,Matsunaga2014,Tsuchiya2018,Tsuji2015, Lancaster-Blundell}:
\begin{equation}
  H = 2 \sum_\kvec 	\begin{pmatrix}
	\xi_k & -\Delta \\
	-\Delta^* & -\xi_k
	\end{pmatrix}
	=
  	2 \sum_\kvec { \bf B}_\kvec \cdot {\bf \Sigma}_\kvec
	=
  	2 \sum_\kvec \mathcal{H}_k,
	\label{eq:H_spin}
\end{equation} 
where ${\bf B}_\kvec = \{- \operatorname{Re}(\Delta) , -\operatorname{Im}(\Delta), \xi_k \}$ is a pseudomagnetic field and ${\bf \Sigma}_\kvec = \{\tau_{x,\kvec} ,\tau_{y,\kvec}, \tau_{z,\kvec} \}$ is the Pauli operator vector \cite{Tsuchiya2018,Tsuji2015,SM}.

Even though $\mathcal{H}_k$ is time dependent, it can be diagonalized as in the standard homogeneous case \cite{deGennes} by introducing the usual (now time-dependent) operators that creates and annihilates the excitations $\gamma_{\kvec \uparrow} = u_k(t) a_{\kvec \uparrow} - v_k(t) a^\dagger_{-\kvec \downarrow}$ and $\gamma^\dagger_{-\kvec \downarrow} = v^*_k(t) a_{\kvec \uparrow} + u^*_k(t) a^\dagger_{-\kvec \downarrow}$ \cite{deGennes,schrieffer1999theory, SM}. The eigenvalues are  $\pm \epsilon_k(t) = \pm \sqrt{\xi_k^2(t) + |\Delta|^2 }$ and the ground and the excited states, expressed in in the original $\{ a_{\kvec \uparrow}, a^\dagger_{-\kvec \downarrow}\}$ basis, are $\ket{\psi_{k,-}(t)} = \{ v_k(t), u_k(t) \}$ and $\ket{\psi_{k,+}(t)} = \{ u_k^*(t) , -v_k^*(t) \}$ respectively, with
$u_k =  \frac{1}{\sqrt{2}} \sqrt{1 + \frac{\xi_k}{\epsilon_k}} e^{-i \chi/2}$ and 
$v_k = \frac{1}{\sqrt{2}} \sqrt{1 - \frac{\xi_k}{\epsilon_k}} e^{i \chi/2}$  \cite{SM}.

\begin{figure*}
    \begin{center}
    \includegraphics[scale=.7]{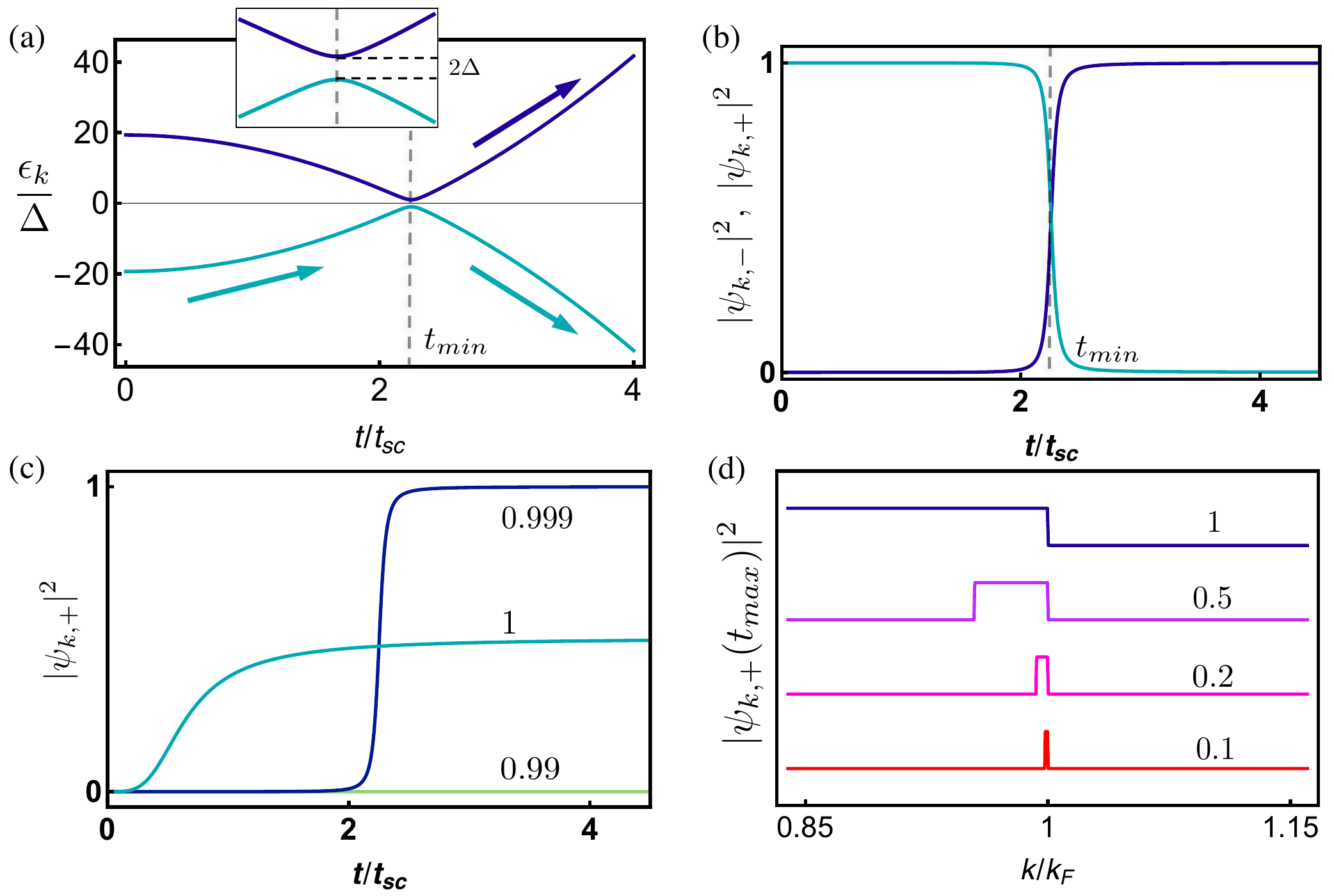}
       \end{center}
    \caption{(a) Spectrum of the superconductor as a function of time $t$ .The inset shows a zoom of the minimal gap region at $t = t_{min}$.
	(b) Ground (cyan) and excited state (blue) populations as a function of time.
	(c) Excited state population as a function of time for initial momentum $k/k_F = 0.99,~0.999$ and $1$.
	In the panels (a) and (b) numerical simulations are performed setting initial momentum $k/k_F = 0.999$ and in (a), (b) and (c) setting $E_f/E_C = 0.2$.
	(d) Excited state final population (at $t=t_{max}$) calculated as a function of  initial momentum $k/k_F$ for selected electric fields  values, $E_f/E_C = 0.1, 0.2, 0.5, 1$. 
	The curves are shifted for presentation purpose.
	Simulations are performed setting $\mu = 1~$eV, $\Delta = 100~\mu$eV and $L=2~$nm and $z/L=0.5$.
     }  
    \label{fig:fig2}
\end{figure*} 

If $\mathcal{U}_k(t)$ is the time-dependent unitary transformation that diagonalizes $\mathcal{H}_k$, namely $\mathcal{U}_k^\dagger \mathcal{H}_k \mathcal{U}_k = \mathcal{H}_{D,k}$, the dynamics is determined by the Schr\"{o}dinger equation
\begin{equation}
i \hbar \partial_t \ket{\phi_k(z)} = (\mathcal{H}_{D,k} - i \hbar~ \mathcal{U}_k^\dagger \partial_t \mathcal{U}_k) \ket{\phi_k(z)}.
\label{eq:schroedinger}
\end{equation}
The contribution $\mathcal{U}_k^\dagger \partial_t \mathcal{U}_k$ induces the transition between eigenstates of $\mathcal{H}_{D,k}$ that are associated to the excitation of the ground state and the creation of quasiparticles.
There is a crucial difference between these kind of excitations  and the well known thermal ones.
The off-diagonal terms in the operator $\mathcal{U}_k \partial_t \mathcal{U}_k^\dagger$ are associated to $ \gamma^\dagger_{\kvec \uparrow} \gamma^\dagger_{-\kvec \downarrow}$ and $ \gamma_{\kvec \uparrow} \gamma_{-\kvec \downarrow}$ \cite{SM}.
Therefore, $\mathcal{U}_k^\dagger \partial_t \mathcal{U}_k$ creates or annihilates simultaneously {\it two quasiparticles} with $(\kvec, \uparrow)$ and $(-\kvec, \downarrow)$.
This is different from the creation of a single quasiparticle through the operators $\gamma^\dagger_{\kvec \uparrow}$ and   $\gamma^\dagger_{-\kvec \downarrow}$ extensively discussed in textbooks \cite{deGennes,tinkham2012introduction}.
While a single excitation destroys a Cooper pair, double excitations preserve the coherent interaction of the pair and, in this sense, they still possess  superconducting properties, as discussed below \cite{Leggett_QuantumLiquids, schrieffer1999theory}.

The pairing potential in Eqs. (\ref{eq:H_spin}) is $\Delta = \sum_k \Delta_k$, and is calculated self-consistently during the dynamics \cite{SM}.
In the numerical simulations, we set $\mu = 1~$eV and the initial pairing potential to $\Delta_0 = 100~\mu$eV \cite{DeSimoni2019,DeSimoniNatNano2018, PaolucciField-Effect,PaolucciNanoLett2018,PaolucciPhysRevAppl2019}. Moreover we assumed the working temperature to be much smaller than superconducting critical temperature $T_C$, eventually neglecting thermal excitations.
For later convenience, we introduce two reference scales: a time scale $t_{sc} = \hbar/\mu$ and a characteristic electric field $E_C = 5 \times 10^8~$V$/$m.

We suppose that a constant electric field is applied at time $t_{in} = 0$, and the dynamical transients are negligible.
The numerical simulations are performed in a finite time interval $0<t<t_{max}$, where $t_{max} = mL/(\hbar ~k)$ is the time needed for a particle of mass $m$ and momentum $k$ to move from one side to the opposite of a sample of thickness $L$ [see Figure \ref{fig:fig1} (a)]. This sets the time scale for the numerical simulations.
In the latter, we set $L=2$~nm \cite{Liu_FeSe_2012, Wang_2012}, so that we can assume a complete penetration of the electric field \cite{Piatti2017}.

The spectrum of the Hamiltonian (\ref{eq:H_spin}) as a function of time is plotted in Fig. \ref{fig:fig2}(a).
The minimum gap $2 \Delta$ is reached when the kinetic energy $\xi_k$ in Eq. (\ref{eq:H_eff_a}) vanishes \cite{SM}, namely for 
\begin{equation}
    t_{min} = \frac{\sqrt{2\mu  m}  \sqrt{1-(k/k_F)^2  }}{e E_f }.
    \label{eq:t_min}
\end{equation}

The dynamics of the populations of the ground and excited state $|\psi_{k,-}|^2$ and $|\psi_{k,+}|^2$ is shown in Fig. \ref{fig:fig2}(b) for a fixed $k/k_F$ and $E_f/E_C$, and presents a clear signature of the SSSE.
The $k$-th mode undergoes a sudden transition to the excited state close to the minimal energy gap. This corresponds to the superconducting Sauter-Schwinger effect, and  to the creation of \emph {two} excited quasiparticles, as discussed above.

The dynamics changes considerably for different initial particle momenta, as shown in Fig.  \ref{fig:fig2}(c).
Away from the Fermi momentum, there is no quasi-particle excitation but moving closer to $k_F$ the system is completely excited.
In a small window very close to $k_F$ the system is only partially excited.

A more complete picture can be inferred from Fig. \ref{fig:fig2}(d) where the final population of the excited state $|\psi_{k,+}(t_{max})|^2$ is shown as a function of the momentum for different normalized electric fields $E_f/E_C$.
For small electric field ($E_f/E_C =0.1$), only a small fraction of the particles around $k_F$ are excited.
By increasing the electric field strength, the excited population fraction increases up to a complete excitation for any $k\leq k_F$ and $E_f/E_C =1$.
These results suggest that the electric field at which the excitations are produced is indeed close to $E_C$ and is remarkably similar to the one used in several recent experiments \cite{DeSimoni2019,DeSimoniNatNano2018, PaolucciField-Effect,PaolucciNanoLett2018,PaolucciPhysRevAppl2019}.

The values of the critical field can be understood treating the SSSE as a Landau-Zener transition (see \cite{SM} for details and \cite{landau1932, zener1932, Shevchenko2010, Cohen2008}).
Imposing the condition that the minimal energy gap is reached earlier than $t_{max}$, i.e., $t_{min} \leq t_{max}$, the electric field needed to produce the excitations: is $E_f/E_C = (2 \mu )/(e E_C L)$ \cite{SM}. For $L=2~$nm, we obtain $E_f = 2 E_C$, namely close to $E_C$, as discussed before.

\begin{figure}
    \begin{center}
    \includegraphics[scale=.70]{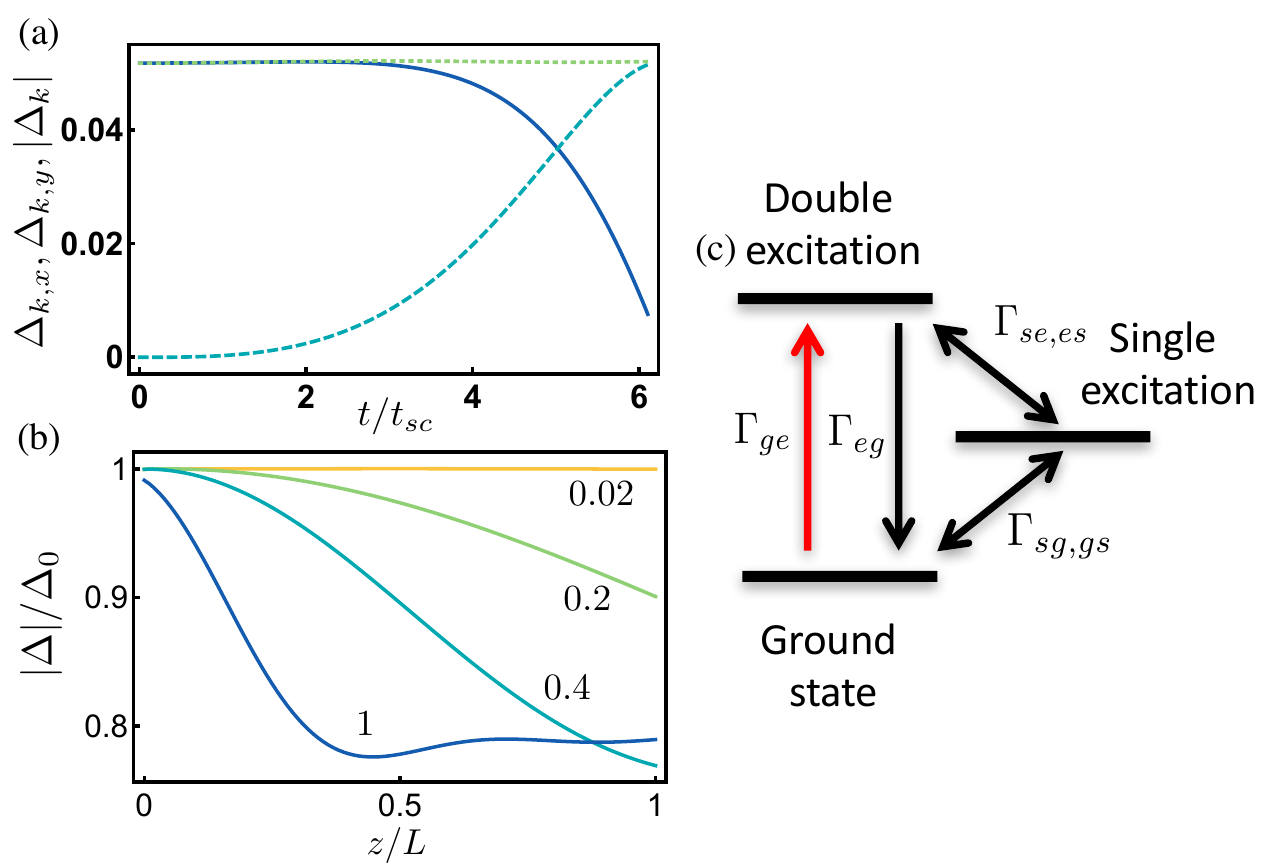}
       \end{center}
    \caption{(a)  Time evolution of $\Delta_{k,x}$ (solid blue), $\Delta_{k,y}$ (dashed cyan) and $|\Delta_k|$ (dotted green) for initial momentum $k/k_F = 0.999$ and $E_f/E_C = 1$. 
   (b) The normalized order parameter $|\Delta|/\Delta_0$ as a function of $z/L$, and for different electric fields $E_f/E_C = 0.02, 0.2, 0.4, 1$.
     (c) The transition scheme associated to the dissipative dynamics of a generic momentum mode.
     The transition rates between the ground and the excited states are denoted with $\Gamma_{ge}$ and $\Gamma_{eg}$ while the ones due to the destruction of the Cooper pairs are $\Gamma_{se}$, $\Gamma_{es}$, $\Gamma_{sg}$ and $\Gamma_{gs}$.
       }  
    \label{fig:fig3V1}
\end{figure}

We stress once again that the double excitation generated by the electric field are deeply different from the single excitation due, for example, to thermal effects.
The latter is related to breaking of a Cooper pair, and leads to an emptying of the pairing potential.
More formally, if a single $k$ excitation is produced, the state $\ket{\psi_{\kvec \uparrow}} = \gamma^\dagger_{\kvec \uparrow} \ketBCS$ (where $\ketBCS = \Pi_k \ket{\psi_{k,-}}$ is the BCS ground state) does not give contribution to the pairing potential, since 
$\Delta_k = \matrixel{\psi_{\kvec\uparrow}}{a_{\kvec \uparrow } a_{-\kvec \downarrow}}{\psi_{\kvec\uparrow}} = 0$.

By contrast, the two excitations produced by the electric field are still correlated, and contribute to the pairing potential but with \emph{opposite} sign with respect to their contribution to the ground state.
More precisely, $\Delta_{GS,k} = \braBCS a_{\kvec \uparrow } a_{-\kvec \downarrow}\ketBCS = u_k v^*_k$ and
$\Delta_{EX,k} = \matrixel{\psi_{k,+}}{a_{\kvec \uparrow } a_{-\kvec \downarrow}}{\psi_{k,+}} = - u_k v^*_k$ \cite{SM, Leggett_QuantumLiquids}.
This has two major implications.
A fully excited state would be still superconducting, preserving all the spectral properties of the ground state, since $\Delta_{EX} = \sum_k \Delta_{EX,k} = - \Delta_{GS}$.
The additional minus sign accumulated in $\Delta_{EX,k}$ can be interpreted as $\pi$-shift of the superconducting phase that, for the excited state, becomes $e^{i (\chi+\pi) }$.
Even though this phenomenon has been discussed in a abstract way in a few books \cite{schrieffer1999theory, Leggett_QuantumLiquids}, this is, to our knowledge, the first time that these elusive correlated excitations could be related to macroscopic effects, and indirectly observed. 
If the $k$-th mode of the ground state is excited, its negative contribution to $\Delta$ can decrease the pairing potential.  We expect a weakening of superconductivity related to these interferences, even though it is worth to mention that in a more general picture the $k$-th mode could be in a coherent superposition of ground and excited states.

In the Anderson pseudo-spin formalism,  the order parameter for the $k$-th  mode is $\Delta_k = \Delta_{k,x} + i  \Delta_{k,y} = \average{\tau_{x,\kvec}} + i \average{\tau_{y,\kvec}}$ where the average $\average{}$ is calculated with state obtained by the dynamical evolution \cite{Anderson1958,Matsunaga2014,Tsuchiya2018,Tsuji2015}.
The numerical calculation displayed in Fig. \ref{fig:fig3V1}(a) shows that while $|\Delta_k|$ is constant, $\Delta_{k,x}$ and  $\Delta_{k,y}$ change in time signaling an accumulated phase.
The pairing potential at $t=t_{max}$ is shown in Fig. \ref{fig:fig3V1}(b) for different electric field values. As the electric field increases the pairing potential is reduced because of the interference effects.
Despite the fact that the environmental effect are not included, this already gives strong indications that the presence of a static electric field  drastically weakens superconductivity.

Interestingly, the SSSE should be associated with non-equilibrium phenomena which can be eventually measured. 
We neglect the momentum scattering leading to a reduction of kinetic energy \cite{giazotto2006opportunities} and discuss the destruction of the excited states with no momentum change, which is the relevant process for the generation of non-equilibrium features (see Fig. \ref{fig:fig3V1}(c) and \cite{SM} for technical details).
Eventually,  reasonably assuming that  the time scales associated with these two processes are well separated allows us to focus only on the second one.

A laser analogy can help to understand how non-equilibrium, i.e., non-thermal, distributions can arise in this context. As in a laser, the electric field acts as an external pumps that excites the ground state directly to the double-excited state. This is unstable and can decay directly to the ground state or through the single excitation state with the Cooper pair destruction. In this way, the standard single excitation states are populated.
The balance between the energy pumping due to the electric field, the dissipation of kinetic energy through momentum scattering and relaxation will eventually lead to a steady state. 
However, while the transitions by and from the single excitation state are thermal, the transition between the ground and double excited state is not. Therefore, as in a laser, the steady state is not related to a thermal or non-equilibrium distribution  \cite{SM}.

The fact that the single-excitation state are, in general, non-thermal, opens the way to a direct measure of the SSSE through tunnel spectroscopy \cite{tinkham2012introduction}.
When two superconductors with different gap, i.e., $\Delta_1$ and $\Delta_2$ with $\Delta_1 < \Delta_2$, are connected with a tunnel junction and subject to a voltage $V$, at finite temperature $T$ the $I-V$ curve shows a resonance current peak at $eV = |\Delta_1-\Delta_2|$ \cite{tinkham2012introduction} due to the presence of quasi-particle. 
If $\Delta_1, \Delta_2 \ll k_B T$ and $E_f=0$, the thermal excitation should be negligible and no current peak at $ |\Delta_1-\Delta_2|$ should be observed.
If we apply a static electric to the superconductor with gap $\Delta_1$, according to the SSSE model with dissipation discussed above, the increase in $E_f$ should generate double and single excitations and this should results in a current resonance at $ |\Delta_1-\Delta_2|$.
The presence of this resonance peak at low temperature and the corresponding excess of (non-thermal) quasiparticles created by the electric field would be the first direct observation of the SSSE. 

The SSSE and its manifestations should be observable with currently available laboratory techniques.
Indeed, the described phenomenology is compatible with recent experiments performed on superconducting wires and nanobridges subject to strong electric fields.
The observed weakening of superconductivity, the invariance of the electric field direction \cite{DeSimoni2019,DeSimoniNatNano2018,PaolucciField-Effect, PaolucciNanoLett2018, PaolucciPhysRevAppl2019, rocci2020gate}, and the creation of non-thermal switching supercurrent distributions \cite{Puglia2020} are compatible with and could be the manifestation of the creation of exotic paired excitations.

Surprisingly, our simple model predicts, with no fitting parameters, that the Sauter-Schwinger effect in superconductors should manifest itself in the presence of an electric field of order $E_C \sim 10^8~$V$/$m, which is in striking agreement with the one used in these experiments \cite{DeSimoni2019,DeSimoniNatNano2018,PaolucciField-Effect, PaolucciNanoLett2018, PaolucciPhysRevAppl2019, alegria2020high, ritter2020superconducting}.


\section{Acknowledgements}
The authors acknowledge S. Paraoanu and  N. Magnoli for fruitful discussions. 
PS and AA acknowledge financial support from INFN. AA has been
partially supported by the INFN Scientific Initiative SFT: “Statistical Field Theory,
Low-Dimensional Systems, Integrable Models and Applications”.
FG acknowledges the Horizon 2020 innovation programme under Grant Agreement No. 800923-SUPERTED for partial financial support.


\pagebreak
\widetext
\begin{center}
\textbf{\large Supplemental Information}
\end{center}
\setcounter{equation}{0}
\setcounter{figure}{0}
\setcounter{table}{0}
\setcounter{page}{1}
\makeatletter
\renewcommand{\theequation}{S\arabic{equation}}
\renewcommand{\thefigure}{S\arabic{figure}}
\renewcommand{\bibnumfmt}[1]{[S#1]}
\renewcommand{\citenumfont}[1]{S#1}

\section{General framework}

The general effective Hamiltonian describing a standard superconductor is \cite{deGennes}
\begin{eqnarray}
 H_{eff} = \int d\rvec
 &\Big\{& 
	\sum_\alpha \Big [ \Psi^\dagger (\alpha \rvec) H_e( \rvec)  \Psi (\alpha \rvec) + \Delta(\rvec) \Psi^\dagger ( \rvec \uparrow)  \Psi^\dagger ( \rvec \downarrow) \nonumber \\
	&& + \Delta^*(\rvec) \Psi (\rvec \downarrow)  \Psi ( \rvec \uparrow)
 \Big\}
 \label{Seq:BCS_H}
\end{eqnarray}
where $\alpha$ is the spin index, $\Psi$ is the fermionic field satisfying the usual anti-commutation rules and, with $V$ set as a coupling energy,  \cite{deGennes}
\begin{eqnarray}
  \Delta(\rvec) &=& - V \average{ \Psi ( \rvec \downarrow )  \Psi ( \rvec \uparrow)}  =  V \average{ \Psi ( \rvec \uparrow)  \Psi ( \rvec \downarrow)} 
  \label{Seq:Delta_def}
\end{eqnarray}
is the self-consistent pair potential.

The single particle Hamiltonian operator is rescaled over the Fermi energy (chemical potential) $\mu$ and it reads
\begin{equation}
	H_e( \rvec) = \frac{1}{2 m} \Big( -i \hbar \nabla - \frac{e}{c} {\bf A} \Big)^2 + U_0(\rvec) -\mu
\end{equation} 
where ${\bf A}$ is the electromagnetic vector potential and $U_0(\rvec)$ is a scalar potential independent on the particle spin.
With this notation we have included the Hartree-Fock potential $U(\rvec) =- V \average{ \Psi^\dagger( \rvec \uparrow)  \Psi ( \rvec \uparrow)}$ in the redefinition of $\mu$ \cite{deGennes}.

We consider thin superconducting films or wires and with limited screening so that the electric field penetrates the superconductor and it is constant inside it.
Alternatively, this model can describe the effect of the electric field on the edge of a metallic superconductor. The electric field $E_f$ is applied to a superconductor along the, say, $z$ direction; i.e., the electric field vector is $\Efvec = \{ 0, 0,  E_f\}$.
Under these hypothesis, we have ${\bf A}=0$ and $U_0(\rvec) = e E_f z$.
However, by a gauge transformation, we can set $U_0(\rvec) = 0$ and ${\bf A}= \{ 0, 0, -c E_f t\}$.

We expand the fermionic fields in Eq. (\ref{Seq:BCS_H}) as
\begin{eqnarray} 
 	\Psi ( \rvec \alpha) &=& \sum_\kvec e^{i \kvec\cdot \rvec} a_{\kvec\alpha} \nonumber \\
	 \Psi^\dagger ( \rvec \alpha) &=& \sum_k e^{-i \kvec\cdot \rvec} a^\dagger_{\kvec\alpha}.
	 \label{Seq:Psi}
\end{eqnarray}
By performing the spatial integration we arrive at \cite{Tsuchiya2018,Tsuji2015, Lancaster-Blundell}
\begin{eqnarray}
 H_{eff}
 &=&  \sum_\kvec \Big \{
 h_{\kvec_- }(t) (a^\dagger_{\kvec \uparrow} a_{\kvec \uparrow} + a^\dagger_{\kvec \downarrow} a_{\kvec \downarrow}) - \Delta a^\dagger_{\kvec \uparrow} a^\dagger_{-\kvec \downarrow} - \Delta^* a_{\kvec \uparrow } a_{-\kvec \downarrow}
 \Big \} 
 \label{Seq:H_eff_a}
\end{eqnarray}
where we have put  $k^2_\perp = k^2_x+k^2_y$. The pairing potential reads 
$\Delta = V \sum_\kvec \average{ a_{\kvec \uparrow } a_{-\kvec \downarrow} }$
\cite{Tsuchiya2018,Tsuji2015, Lancaster-Blundell}.

It is convenient to simplify the notation but, at the same time, keep track of the presence of the vector potential $ {\bf A}$.
For this reason, we introduce the kinetic energy
$ h_{\kvec_- }(t) = h_{\kvec - \frac{e}{\hbar c} {\bf A}}(t) =\frac{1}{2m} \Big[ \hbar^2 k^2_\perp + (\hbar k_z + e E_f t)^2 \Big] -\mu$.

In the sum in Eq. (\ref{Seq:H_eff_a})  both positive and negative $\kvec$ contributions are present.
We can separate the negative terms like $h_{-\kvec_- }  a^\dagger_{-\kvec \uparrow} a_{-\kvec \uparrow}$.
We have 
\begin{equation}
  h_{-\kvec_- } = h_{-\kvec - \frac{e}{\hbar c} {\bf A}}(t) = \frac{1}{2m} \Big[ \hbar^2 k^2_\perp + (\hbar k_z - e E_f t)^2 \Big] -\mu = h_{\kvec_+ } 
  \label{Seq:minus_k}
\end{equation}
Thus, formally reversing the momentum is equivalent to change the charge to the particle.

The superconductor pair potential can be written as $\Delta = |\Delta| e^{i \chi}$ where $\chi$ is the superconduting phase.
It is related to the gauge-invariant scalar $\phi$ and vector ${\bf{A}}$ potentials by the equations \cite{Kopnin2002}
\begin{eqnarray}
 {\bf A} &=&  {\bf \mathcal{A}} - \frac{\hbar c}{2 e} \nabla \chi \nonumber \\
 \phi &=& \mathcal{V} + \frac{\hbar}{2 e}  \frac{ \partial \chi}{\partial t}.
\end{eqnarray}
These are related to the physical electric ${\bf E}$ and magnetic field ${\bf h}$ by the relations \cite{Kopnin2002}
\begin{eqnarray}
 {\bf E} &=&  - \frac{1}{c} \frac{\partial {\bf \mathcal{A}}}{\partial t} - \nabla \mathcal{V}  \nonumber \\
 {\bf h}&=&  \nabla \times {\bf \mathcal{A}}.
\end{eqnarray}

By setting $\phi =0$ and ${\bf \mathcal{A}}=0$, i.e., no magnetic field, we obtain 
\begin{equation}
	\chi = \frac{2 e}{\hbar} E_f~t~z
	\label{Seq:super_phase}
\end{equation}
and ${\bf A}= \{ 0, 0, - c E_f t\}$ as above.
Therefore, the superconducting phase, the pairing potential (\ref{Seq:Delta_def}) and the Hamiltonian (\ref{Seq:H_eff_a}) depends on the spatial coordinate $z$.

This gauge choice allows us to deal with a homogeneous problem where the spatial dependence has vanished in Eq. (\ref{Seq:H_eff_a}).
This is a great simplification because allows to use the standard approach and techniques to describe the superconducting state and dynamics.
The price to pay for this simplification is to deal with a time-dependent Hamiltonian so that we are forced to solve the time-dependent dynamics.
Because the problem is homogeneous, only the $(\kvec, \uparrow)$ and $(-\kvec, \downarrow)$ are coupled. This makes the problem easily solvable both numerically and analytically.

We can collect the terms in Eq. (\ref{Seq:H_eff_a}) separating the $\kvec$ and the $-\kvec$ contributions.
By using the state ${\bf \Phi} =\{ a_{\kvec \uparrow} , a^\dagger_{-\kvec \downarrow}\}$, the relation $h_{-\kvec_- } = h_{\kvec_+ }$ and the anti-commutation rules for fermionic operators $a^\dagger_{\kvec \alpha}$ and $a_{\kvec \alpha}$, we can rewrite Eq. (\ref{Seq:H_eff_a}) in matrix form as \cite{Tsuchiya2018,Tsuji2015,Lancaster-Blundell}
\begin{equation}
  H_{eff} = 2 \sum_\kvec 	\begin{pmatrix}
	\xi_k & -\Delta \\
	-\Delta^* & -\xi_k
	\end{pmatrix}
	=
  	2 \sum_\kvec { \bf B}_k \cdot {\bf \Sigma}_k
	=
  	2 \sum_\kvec \mathcal{H}_k
	\label{Seq:H_spin}
\end{equation} 
where 
\begin{equation}
 	\xi_k = \frac{h_{\kvec_-}+h_{\kvec_+}}{2} =\frac{\hbar ^2 k^2}{2 m}+ \frac{e^2 E_f^2 t^2}{2 m}-\mu,
	\label{Seq:xi_k}
\end{equation}
${\bf B}_k = \{- \operatorname{Re}(\Delta) , -\operatorname{Im}(\Delta), \xi_k \}$ is a pseudo-magnetic field and ${\bf \Sigma}_k = \{\tau_{x,k} ,\tau_{y, k}, \tau_{z, k} \}$.
This is nothing but the the Anderson pseudospin approach \cite{Anderson1958,Matsunaga2014,Tsuchiya2018,Tsuji2015}.

\section{Quasi-particle creation: the superconductor Schwinger effect}
\label{sec:schwinger_effect}

To highlight the Schwinger effect and the creation of quasi-particles, it is convenient to use the representation that diagonalizes (\ref{Seq:H_spin}).
This is the approach used in an alternative derivation of the original Schwinger effect in quantum electrodynamics  in Ref. \cite{Cohen2008}.

The operator $\mathcal{H}_k$ has the same form of the standard homogeneous case and can be analytically diagonalized \cite{deGennes}.
The eigenvalues are  $\pm \epsilon_k = \pm \sqrt{\xi_k^2 + |\Delta|^2 }$ and the ground and the excited states are, in the original $\{ a_{\kvec \uparrow}, a^\dagger_{-\kvec \downarrow}\}$ basis, $\ket{\psi_{k,-}(t)} = \{ v_k(t), u_k(t) \}$ and $\ket{\psi_{k,+}(t)} = \{ u_k^*(t) , -v_k^*(t) \}$, respectively, with
\begin{eqnarray}
 u_k(t) &=&  \frac{1}{\sqrt{2}} \sqrt{1 + \frac{\xi_k(t)}{\epsilon_k(t)}} e^{-i \chi(t)/2}\nonumber \\ 
  v_k(t) &=& \frac{1}{\sqrt{2}} \sqrt{1 - \frac{\xi_k(t)}{\epsilon_k(t)}} e^{i \chi(t)/2}.
  \label{Seq:uv_BdG}
\end{eqnarray}

Be $\mathcal{U}_k(t)$ the diagonalizing operator such that $\mathcal{U}_k^\dagger \mathcal{H}_k \mathcal{U}_k = \mathcal{H}_{D,k}$.
Since $\mathcal{U}_k$ is time dependent, the dynamics is determined by the Schroedinger equation
\begin{equation}
i \hbar \partial_t \ket{\psi_k(z)} = (\mathcal{H}_{D,k} - i \hbar~ \mathcal{U}_k^\dagger \partial_t \mathcal{U}_k) \ket{\psi_k(z)}.
\label{Seq:schroedinger}
\end{equation}
The contribution $\mathcal{U}_k^\dagger \partial_t \mathcal{U}_k$ derives from the fact that the Hamiltonian is time-dependent and induces the transition between eigenstates of $\mathcal{H}_{D,k}$.
Notice that Eq. (\ref{Seq:schroedinger}) depends on $z$. Thus, it gives us the dynamics of the $k$-th mode in position $z$.

\subsection{Double excitations}

The unitary operators $\mathcal{U}_k(t)$ and $\mathcal{U}^\dagger_k(t)$ can be written as
\begin{eqnarray}
 \mathcal{U}_k &=& 	
 	\begin{pmatrix}
	u^*_k & v_k \\
	-v^*_k & u_k
	\end{pmatrix} \nonumber \\
	 \mathcal{U}^\dagger_k &=& 	
 	\begin{pmatrix}
	u_k & -v_k \\
	v^*_k & u^*_k
	\end{pmatrix}
	\label{Seq:U_k_matrix}
\end{eqnarray}
with $u_k$ and $v_k$ as in Eq. (\ref{Seq:uv_BdG}).
This leads to the transformation \cite{deGennes,tinkham2012introduction}
\begin{eqnarray}
	\gamma_{\kvec \uparrow} &=& 	u_k a_{\kvec \uparrow} - v_k a^\dagger_{-\kvec \downarrow} \nonumber \\
	\gamma^\dagger_{-\kvec \downarrow} &=& v^*_k a_{\kvec \uparrow} + u^*_k a^\dagger_{-\kvec \downarrow} \nonumber \\
	\gamma^\dagger_{\kvec \uparrow} &=& 	u^*_k a^\dagger_{\kvec \uparrow} - v^*_k a_{-\kvec \downarrow} \nonumber \\
	\gamma_{-\kvec \downarrow} &=& v_k a^\dagger_{\kvec \uparrow} + u_k a_{-\kvec \downarrow}.
	\label{Seq:gamma_def}
\end{eqnarray}
These are the creation-annihilation operators for a quasiparticle that is a superposition of electron and hole \cite{deGennes,tinkham2012introduction, Lancaster-Blundell}.

In this representation the diagonal element of $\mathcal{H}_{D,k}$ are associated to the $ \gamma^\dagger_{\kvec \uparrow} \gamma_{\kvec \uparrow}$ and  $\gamma^\dagger_{-\kvec \downarrow} \gamma_{-\kvec \downarrow}$.
On the contrary, the $\mathcal{U}_k^\dagger \partial_t \mathcal{U}_k$ off-diagonal terms are associated to $ \gamma^\dagger_{\kvec \uparrow} \gamma^\dagger_{-\kvec \downarrow}$ and $ \gamma_{\kvec \uparrow} \gamma_{-\kvec \downarrow}$ and, therefore, create or annihilate simultaneously {\it two quasiparticles} with $(\kvec, \uparrow)$ and $(-\kvec, \downarrow)$. These are what in the original BCS paper are called "real" excited pairs \cite{schrieffer1999theory, BCS_Micro_1957} and Leggett associates with the natural excitation in the $(\kvec, \uparrow)$ and $(-\kvec, \downarrow)$ space since they are still part of the condensate
\cite{Leggett_QuantumLiquids}.
They must be distinguished by the conventional "Bogoliubov quasiparticles" discussed in literature \cite{deGennes, tinkham2012introduction} that are related to the destruction of a Cooper pair.

Using the Anderson pseudo-spin formalism is easy to understand the nature of the excited state $\ket{\psi_{k,+}}$.
The pairing potential is defined as $\Delta = \sum_k \Delta_k$ with $\Delta_k = V \average{ a_{\kvec \uparrow } a_{-\kvec \downarrow} }$.
For the ground state $\ket{\psi_{k,-}}$, we obtain $\Delta_k =V u_k v^*_k$ as expected \cite{deGennes, tinkham2012introduction}.
For the excited state, we have $\Delta_k = - V u_k(t) v^*_k(t)$ \cite{Leggett_QuantumLiquids}.
This can be seen as an additional phase factor $e^{i \pi}$ or, alternatively, a $\pi$ shift in the superconducting phase associated to $\Delta_k$ due to the ground-excited transition.

We conclude that the excited $k$ states preserve the superconductive feature.
While the single excitation states (Bogoliubov quasiparticles) are associated to a vanishing coherence factor, i.e., $\average{ a_{\kvec \uparrow } a_{-\kvec \downarrow} }= 0$, the double excitation states $\ket{\psi_{k,+}}$ are associated to the same coherent factor with a minus sign.
This means that a fully excited state, i.e., with all the $k$ modes excited, would have the same pairing potential and the same gap.
In this sense, the excited state is still superconducting or, in Leggett's words, is still “part of” the condensate \cite{Leggett_QuantumLiquids}.

The fact that the excited pair state are still superconducting have important implications.
The single excitation Bogoliubov quasiparticles are obtained by the destruction of a Cooper pair and vanishing coherence factor.
This leads to the emptying of the condensate since the excited modes do not contribute to the pairing potential $\Delta = \sum_k \Delta_k$.

On the contrary, the state generated by the SSE, i.e., the superposition of ground and excited state, is still superconductive. 
But since the dynamics of the $k$ modes is different, they accumulate a different phase factors.
The coherence factor $\Delta_k$ is in general a complex number with a time dependent phase.
The different phases can generate "interference effects" in the sum $\Delta = \sum_k \Delta_k$ effectively leading to a suppression of the pairing potential and the superconductivity as shown in Fig. 
\ref{fig:fig3V1}b 
of the main text.
We stress once more that the mechanism at the basis of the destruction of superconductivity in presence of an electric field is completely distinguished and new with respect to the usual thermal one.

An important remark must be done.
As discussed above, a fully excited state would be similar in every aspect to the ground state.
However, it turns out that the current associated to the ground and the excited state is the same.
Therefore, the current is not a good observable to distinguish the SSE.

\subsection{Numerical simulations}
\label{sec:numerical_simulations}

The numerical simulations are performed using the self-consistent relation for the pairing potential (\ref{Seq:Delta_def}).
The solution scheme is presented for a given spatial point $z$ and the procedure can be iterated for different $z$ to obtain the spatial behaviour of the main superconductor quantities.

The $k$ involved in the superconductivity are the ones with $k_F - k_D \leq k \leq k_F+ k_D$ where $k_F$ and $k_D$ are the Fermi and the Debye momentum, respectively \cite{deGennes}.
As a reference, we have taken $\mu =\hbar^2 k_F^2/(2 m)=1~$eV ($m$ is the electron mass) and a Debye temperature of $300~$K \cite{deGennes} corresponding to a normalized momentum $k_D/k_F = 0.16$ that are standard values of BCS superconductors. 
The momentum space is divided in $\Delta k$ intervals and the dynamics is computed for any $k_n = k_F - k_D + n~ \Delta k$ (with $n$ integer) in the useful interval.

The state of the system is initialized in the ground state.
Thus, all the $k$-th modes are initially in the ground state $\ket{\psi_{k,-}(0)}$ of the initial Hamiltonian (\ref{Seq:H_spin}).
For all the relevant $k$, the numerical code calculates the solution $\ket{\psi_k(\Delta t)}$ of a discretized version of the Schroedinger equation (\ref{Seq:schroedinger}) for small time increment $\Delta t$.
Then, with all the $\ket{\psi_k(\Delta t)}$, it calculates the new pairing potential $\Delta(\Delta t) = \sum_k \Delta_k$ with 
$\Delta_k = \matrixel{\psi_k(\Delta t)}{a_{\kvec \uparrow} a_{-\kvec \downarrow}}{\psi_k(\Delta t)}$.
The updated pairing potential is inserted in the Schroedinger equation for the calculation of the following time evolution.

All the numerical results presented are calculated in a self-consistent way but it turns out that the differences with the non-self-consistent case, i.e., with $\Delta$ constant in time, are small.

\section{Landau-Zener transition and relevant parameters}
\label{sec:LZ}

The overall physical features of the dynamics can be understood with an analogy simple Landau-Zener problem \cite{landau1932, zener1932, Shevchenko2010}.
We discuss the case in which $|\Delta|$ is constant in time and not calculated with the self-consistent relation. 
This approximation not only allows us understand the main physical features of the dynamics but it turns out to be an excellent approximation of the full self-consistent dynamics.

Let us consider first the case in which $k < k_F$.
At $t=0$ the kinetic energy dominates and $\mathcal{H}_k (0) \approx \xi_k \tau_{z,k}$.
The minimum energy is reached at $t_{min}$ when $\xi_k(t_{min})=0$ and the Hamiltonian $\mathcal{H}_k (t_{min}) = \operatorname{Re}(\Delta) \tau_{x,k} + \operatorname{Im}(\Delta) \tau_{y,k}$. Finally, for $t > t_{min}$ the kinetic energy terms $h_{\kvec_-}(t)$ dominates over $\Delta$ and $\mathcal{H}_k(t) \approx h_{\kvec_-}(t) \tau_{z,k}$.

Thus, we have a Landau-Zener problem \cite{landau1932, zener1932, Shevchenko2010} with $i)$ the typical Hamiltonian changes $\tau_{z,k} \rightarrow \tau_{x,k} \rightarrow \tau_{z,k}$, $ii)$ an avoided crossing at $t_{min}$ with energy gap $2 \Delta$ and $iii)$ the system in the ground state $\mathcal{H}_k (0)$ at the beginning of the evolution.

To have an estimate of the transition probability between instantaneous eigenstates of $\mathcal{H}_k$ (also called adiabatic states), we use the approach discussed in Ref. \cite{Shevchenko2010}.
We set $t= t_{min} + \delta t$ with $\delta t \ll t_{min}$ and expand and linearize $\xi_k(t_{min})$ for small $\delta t$ to obtain the normalized (to $\mu$) {\it energy velocity} close to the minimum gap
\begin{equation}
	V_k = \sqrt{2} e E_0 \sqrt{\left[1-\left(\frac{k}{k_F}\right)^2\right] \frac{\mu}{m}}.
\end{equation}
This would correspond to $\partial \epsilon_k/\partial t$ of the original Landau-Zener model for linear drive \cite{Shevchenko2010}.

The probability to have a transition from the ground to the excited state (so called Landau-Zener probability) with minimum energy gap $2 \Delta$ is \cite{Shevchenko2010} 
\begin{equation}
	P_{LZ} = e^{-\frac{2 \pi  \Delta ^2}{V_k \hbar }}.
	\label{eq:LZ}
\end{equation}
In our case, using the above expression and the scaling parameters, the Landau-Zener coefficient reads
\begin{equation}
 \Gamma_k = \frac{2 \pi  \Delta ^2}{ V_k \hbar } = \frac{1.6 \times 10^{-6}}{\frac{E_f}{E_C} \sqrt{1-\left(\frac{k}{k_F}\right)^2}}.
 \label{eq:LZ_Gamma}
\end{equation}
Therefore,  for (almost) any $k$, we have $\Gamma_k \ll 1$ and $P_{LZ} \approx 1$ and the $k$ modes should undergo a ground-to-excited  state transition (see also discussion below).

The situation is different for initial momentum very close to the Fermi momentum, i.e., $1- k/k_F \approx 10^{-5}$ where the Landau-Zener model is inadequate to describe the dynamics.

In this case, the initial Hamiltonian is $\mathcal{H}_k(0) \approx -\Delta \tau_{x,k}$ and the system is in its ground state $\ket{\psi_{k,-}(0)} = 1/\sqrt{2}~ \{1, 1\}$ (since $u_k \approx v_k \approx 1/\sqrt{2}$).
At the minimum energy gap, $\mathcal{H}_k(t_{min}) \approx -\Delta \tau_{x,k}$. 
Thus, during the first part of evolution $ 0\leq t\leq t_{min}$, $\mathcal{H}_k(t) \propto \tau_{x,k}$.
A key ingredient of the Landau-Zener model, i.e., the first Hamiltonian change $\tau_{z,k} \rightarrow \tau_{x,k} $, is now missing.
Since the system remains in an eigenstate of the Hamiltonian, the first part of the evolution accounts only for a dynamical phase factor and {\it no Landau-Zener transition} between eigenstates occurs.

For $t> t_{min}$, the kinetic energy increases and start dominating so that the Hamiltonian is $\mathcal{H}_k(t) \approx \xi_k  \tau_{z,k}$.
During this evolution, no Landau-Zener transition occurs and the system remains in the ground state of $\tau_{x,k}$.
Decomposing this in the eigenstates of $\tau_{z,k}$ we obtain that the ground and excited states are equally populated as shown in Fig. 
\ref{fig:fig2}c 
of the main text.

The last relevant feature of the dynamics is that for $k \ll k_F$, there is no transition to the excited state (Fig. \ref{fig:fig2}c 
in the main text). This feature can be understood by comparing the times $t_{min}$ and $t_{max}$.
The first one sets the time in which the minimum energy gap is reached and the transition occurs and it is determined by the electric field; the second one sets the time for the dynamics and it is determined by the spatial length $L$ of the system.

We recall the expressions
\begin{equation}
    t_{min} = \frac{\sqrt{2\mu  m}  \sqrt{1-(k/k_F)^2  }}{e E_f }
    \label{Seq:t_min}
\end{equation}
and $t_{max} = mL/(\hbar ~k)$.
The condition $t_{min} = t_{max}$ gives the limit for the transition to occur.
From these, we obtain an equation for the minimum electric field needed to excite the mode $k$ (neglecting the $k$ dependence)
\begin{equation}
 \frac{E_f}{E_C} = \frac{2 \mu }{e E_C L}.
\end{equation}
Below this value the dynamics is not long enough to reach the minimal gap or, rephrasing, the electric field has not enough time to give to the superconductor mode enough energy to undergo the ground-to-excite transition.

For $L=2~$nm, we obtain $E_f = 2 E_C$, namely close to $E_C= 5 \times  10^8~$V/m that is the electric field at which the superconductivity is suppressed in several recent experiments
\cite{DeSimoni2019mesoscopic,DeSimoniNatNano2018,PaolucciField-Effect, PaolucciNanoLett2018, PaolucciPhysRevAppl2019, Puglia2020, rocci2020gate}.

The Landau-Zener approach and in particularly Eqs. (\ref{eq:LZ}) and (\ref{eq:LZ_Gamma}), allow us to understand why the numerical simulations with and without self-consistent calculations of the pairing potential give the same results.

The Landau-Zener coefficient (\ref{eq:LZ_Gamma}) is very small for (almost) any initial momentum.
This is due to the combination of parameter in the system and any physical and realistic energy gap (of the order of $100~\mu$eV) would give practically the same probability transition.
To make an example, we can take the minimal $\Delta = 0.8 \Delta_0$ as obtained in the self-consistent numerical shown in Fig. $3$(c) in the main text. In Eq. (\ref{eq:LZ_Gamma}), the numerator would result in $10^{-6}$ that, for any practical purposes, is not distinguishable from $1.6 \times 10^{-6}$. Therefore, using the self-consistent pairing potential does not change the transition probability in any appreciable way.

\section{Out-of-equilibrium quasi-particles}

It is known that even a constant electric field in a superconductor generates out-of-equilibrium phenomena \cite{kopnin2001theory}.
The fundamental reason for this is that charged particles in the superconductor are accelerated and increase their energy.
As a direct consequence, the effect of the environment must be included in the treatment otherwise the energy increase would inevitably lead to the destruction of superconductivity.

The SSE can naturally lead to non-equilibrium or out-of-equilibrium phenomena but also in this situation we must include some dissipative channel.
The electric field acts as a pump increasing of the superconductor energy while the dissipation tends to reduce it.
The steady state would be reach when the injected and dissipated energy rates are equal.

The energy dissipation can occur in two ways.
The first one is the scattering where a particle with momentum $k$ is scattered into a particle of momentum $k'$ and it partially dissipates its kinetic energy.
The second is the destruction (or creation) of excited states. These are, in general, distinguished dissipation mechanisms.
For example, the destruction of a double excitation state is associated to the decrease in energy of $2 \Delta$ but there is no scattering of momentum since the transition does not change $\kvec$ of the particles, i.e., it remains within in the $(\kvec \uparrow, -\kvec \downarrow)$ space.

Without pretending to be a quantitative description, a simple model can help us to understand how non-equilibrium, i.e., non-thermal, distributions can arise in this contest.
We make the working hypothesis that the two momentum scattering and creation/annihilation time scales are well separated.
This allows us to treat separately the energy process due to the creation/destruction of excited states and the $k$ scattering the dissipated the kinetic energy and focus on the first one.
In fact, the dissipation through a scattering process is fairly well understood \cite{giazotto2006opportunities} while the non-equilibrium features (the most interesting for us) are consequences of the first process.

We use a simple master equation to describe the transition between the state in the $(\kvec \uparrow, -\kvec \downarrow)$ as shown schematically in Fig. 
\ref{fig:fig3V1}c of the main text.
From the ground state $\ket{\psi_{k,-}}$ we can excite a double excited state $\ket{\psi_{k,+}} = \gamma^\dagger_{\kvec \uparrow} \gamma^\dagger_{-\kvec \downarrow} \ket{\psi_{k,-}}$ and, in presence of an environment, the latter can relax back to the ground state.

In addition to these, we must include the possibility to break a Cooper pair and generate a single excitation state $\ket{\psi_{s_1}}= \gamma^\dagger_{\kvec \uparrow} \ket{\psi_{k,-}} = \gamma_{\kvec \uparrow} \ket{\psi_{k,+}}$ and $\ket{\psi_{s_2}} = \gamma^\dagger_{-\kvec \downarrow}\ket{\psi_{k,-}} = \gamma_{-\kvec \downarrow}\ket{\psi_{k,+}}$.
Formally these are obtained with a single destruction ($\gamma$) or creation ($\gamma^\dagger$) operator applied to the excited or ground state, respectively.
Notice that, without the electric field and the double exited state, these are the only ones that are thermally excited and result in a Fermi-Dirac distribution for the quasi-particles (the thermal double excitation is exponentially suppressed).

We denote with $\Gamma_{eg}$ and $\Gamma_{ge}$, the excitation and relaxation rates between the excited and ground state, respectively.
Since the environment does not distinguish between the creation or destruction of a quasiparticle with $(\kvec, \uparrow)$ or $(-\kvec \downarrow)$, the two single excited state $\ket{\psi_{s_1}}$ and $\ket{\psi_{s_2}}$ are indistinguishable. 
It is convenient to describe these transitions in terms of single transition to a single excitation state $\ket{\psi_s}$.
The environment induces transitions from and to the excited states with rates $\Gamma_{es}$ and $\Gamma_{se}$, respectively, and from and to the ground state with rates $\Gamma_{gs}$ and $\Gamma_{sg}$, respectively (see Fig.$3$c of the main text).

For a given mode $\kvec$, the corresponding master equations for the ground  $\mathcal{P}_g$, single excitation  $\mathcal{P}_s$ and excited state populations $\mathcal{P}_e$ read
\begin{eqnarray}
 \partial_t \mathcal{P}_g &=&  - \left(\Gamma _{ge}+2 \Gamma _{gs}\right) \mathcal{P}_g+  \Gamma_{eg} \mathcal{P}_e+  \Gamma
   _{sg}  \mathcal{P}_s \nonumber \\
 \partial_t \mathcal{P}_s &=&  -
   \left(\Gamma _{se}+\Gamma _{sg}\right)\mathcal{P}_s+ 2  \Gamma _{es} \mathcal{P}_e +2  \Gamma _{gs} \mathcal{P}_g \nonumber \\
  \partial_t \mathcal{P}_e &=& - \left(\Gamma _{eg}+2 \Gamma
   _{es}\right) \mathcal{P}_e +  \Gamma _{ge} \mathcal{P}_g + \Gamma _{se} \mathcal{P}_s .
   \label{eq:master_equation}
\end{eqnarray}
By using the normalization condition $\mathcal{P}_e + \mathcal{P}_s + \mathcal{P}_g =1$, we can reduce these to two differential equations for $\mathcal{P}_g$ and $\mathcal{P}_s$
\begin{eqnarray}
 \partial_t \mathcal{P}_g &=&  - \left(\Gamma _{eg}+\Gamma _{ge}+2 \Gamma _{gs}\right) \mathcal{P}_g+  \left(\Gamma _{sg}-\Gamma _{eg}\right) \mathcal{P}_s \nonumber \\
 \partial_t \mathcal{P}_s &=&
   -\left(2 \Gamma _{es}+\Gamma _{se}+\Gamma _{sg}\right)\mathcal{P}_s+ 2 \left( \Gamma _{gs}- \Gamma _{es}\right) \mathcal{P}_g.
   \label{eq:master_equation_reduced}
\end{eqnarray}
The stationary solution is obtained for $\partial_t \mathcal{P}_g = 0$ and $\partial_t \mathcal{P}_s=0$. 
The solution of these algebraic equations expressed in terms of the ratio between the ground and the single excitation state is
\begin{equation}
 \frac{\mathcal{P}_g}{\mathcal{P}_s}\Big|_{steady} = \frac{\Gamma _{eg} \Gamma _{se}+\Gamma _{eg} \Gamma _{sg}+2 \Gamma _{es} \Gamma
   _{sg}}{2 \Gamma _{ge} \Gamma _{es}+2 \Gamma _{eg} \Gamma _{gs}+4 \Gamma _{es} \Gamma _{gs}}.
\end{equation}

We further assume that $i)$ for the {\it thermal transitions}, the relaxation and excitation rates are related by the Boltzmann rules, i.e., $\Gamma _{se} = e^{-\frac{\epsilon_k}{k_B T}} \Gamma _{es}$ and $\Gamma _{gs} = e^{-\frac{\epsilon_k}{k_B T}} \Gamma _{sg}$ ($k_B$ is the Boltzmann constant and $\epsilon_k$ is single excitation energy gap) and $ii)$ $\Gamma _{es} = \Gamma _{sg}$ since the energy gap and the transition amplitudes are the same.
We arrive to 
\begin{equation}
 \frac{\mathcal{P}_g}{\mathcal{P}_s}\Big|_{steady} = \frac{1}{2} \frac{\Gamma _{eg} \Gamma
   _{sg}+2 \Gamma _{sg}^2 +e^{-\frac{\epsilon _k}{k_B T}} \Gamma _{eg} \Gamma _{sg}
   }{   \Gamma _{ge} \Gamma _{sg} +
     e^{-\frac{\epsilon _k}{k_B T}} ( \Gamma _{eg} \Gamma _{sg}+2  \Gamma _{sg}^2)}.
     \label{eq:stationary_solution}
\end{equation}
The rates $\Gamma _{eg}$ and $\Gamma _{ge}$ are not Boltzmann-related and, in particular, $\Gamma _{eg}$ depends only on the electric field effect (if $2 \Delta \ll k_B T$).
Therefore, the above expression cannot be related to a simple decay or thermalization and would result in a non-thermal distribution of quasi-particles.

To have a more direct comparison we can consider the same model without electric field by taking the limit $\Gamma_{ge} \rightarrow 0 $ and $\Gamma_{eg} \rightarrow 0 $ in Eq. (\ref{eq:stationary_solution}).
We obtain the stationary ratio
\begin{equation}
	\frac{\mathcal{P}_g}{\mathcal{P}_s}\Big|_{no~E_f} = \frac{1}{2 e^{-\frac{\epsilon _k}{k_B T}}}
\end{equation} 
that leads, indeed, to a thermal distribution of quasi-particle.

Even more evident is the limit with a few thermal excitations $\epsilon _k \gg k_B T$, in which we have
\begin{eqnarray}
 \frac{\mathcal{P}_g}{\mathcal{P}_s}\Big|_{steady} &\rightarrow& 
 \frac{\Gamma _{eg} +2 \Gamma _{sg}}{2  \Gamma _{ge} } 
 \nonumber \\
  \mathcal{P}_g\Big|_{no~E_f} &\rightarrow& 1 ~ {\rm and }~ \mathcal{P}_s\Big|_{no~E_f} \rightarrow 0
\end{eqnarray}
This, again, shows the differences between the two steady states and the fact that the electric field can generate non-equilibrium distribution.


\section{Interpretation within the gapless superconductivity}
\label{app:gapless_superconductivity}

As discussed in the main text, within our gauge choice, the order parameter acquires a time-dependent phase: 
 $\Delta = |\Delta| e^{i \chi}$ with $\chi = \frac{2 e}{\hbar} E_f t z$.
It could be tempting to make an analogy with the phenomenon of {\it gapless superconductivity} 
\cite{deGennes, tinkham2012introduction, Aronov1981}.
Here we show that this analogy is incorrect, because it describes single quasi-particle excitations while the electric field creates two quasi-particle excitations. 
The latter physical process determines the correct energy gap that does not vanish instead.

For convenience, we review the phenomenon of gapless superconductivity following Ref. \cite{deGennes} and using our notations.
In the presence of an external uniform current flow, the electrons change momentum. The electron-like and the hole-like particle momenta change as $\kvec \rightarrow \kvec + \qvec$ and $-\kvec \rightarrow -\kvec + \qvec$.
The corresponding solutions of the Bogoliubov-de Gennes (BdG) equations can be found assuming the following ansatz: 
$u_k = U_k e^{i (\kvec + \qvec) \cdot \rvec}$ and $v_k = V_k e^{i (\kvec - \qvec) \cdot \rvec}$.
As a consequence of the gauge invariance of the BdG equations \cite{deGennes}, the pairing potential changes as $\Delta = |\Delta| e^{ 2 i \qvec \cdot \rvec} $ and 
the new  BdG equations read
\begin{eqnarray}
 (\tilde{\epsilon}_k - h_{\kvec + \qvec})U_k  - |\Delta|  V_k &=&  0 \ , \nonumber \\ 
  (\tilde{\epsilon}_k + h_{\kvec - \qvec})V_k  - |\Delta|  U_k &=&  0 \ .
  \label{eq:BdG_gapless}
\end{eqnarray}
By solving with respect to $V_k$, we obtain the following gap equation $\tilde{\epsilon}_k^2 - (h_{\kvec + \qvec}-h_{\kvec - \qvec}) \epsilon_k  + h_{\kvec + \qvec}h_{\kvec - \qvec} =0$,
from which one can eventually obtain the energy of the ground and excited states:
\begin{equation}
 \tilde{\epsilon}_k = \frac{h_{\kvec + \qvec}-h_{\kvec - \qvec}}{2} \pm \sqrt{\Big(\frac{h_{\kvec + \qvec}+h_{\kvec - \qvec}}{2}\Big)^2 + |\Delta|^2} .
\label{app_eq:gap_gapless}
\end{equation}

In the limit $|q| \ll |k| \approx k_F$ \cite{deGennes}, $(h_{\kvec + \qvec}+h_{\kvec - \qvec})/2 \approx h_{\kvec}  = \frac{ \hbar^2 k^2}{2m} -\mu$
, $ (h_{\kvec + \qvec}-h_{\kvec - \qvec})/2 = \frac{\hbar^2}{m} \kvec \cdot \qvec$
and we obtain 
\begin{equation}
 \tilde{\epsilon}_k \approx \frac{\hbar^2}{m} \kvec \cdot \qvec \pm \sqrt{\Big(\frac{ \hbar^2 k^2}{2m} -\mu \Big)^2 + |\Delta|^2}.
 \label{Seq:gapless_epsilon_k}
\end{equation} 
Let us focus on the minus sign solution in Eq. (\ref{Seq:gapless_epsilon_k}), restricting to the case in which $\qvec$ and $\kvec$ have the same direction.
At the Fermi momentum $k=k_F$, the kinetic energy vanishes and we have 
\begin{equation}
 \tilde{\epsilon}_k \approx  \frac{\hbar^2}{m} k_F q -  |\Delta|.
 \label{Seq:vanishing_epsilon_k}
\end{equation} 
The minimum {\it ground state energy} is $\tilde{\epsilon}_k = 0$ and it is reached when $q= \frac{m |\Delta| }{ \hbar^2 k_F}$, i.e., when the superfluid velocity $v_s = \hbar q/m$ reaches the value $|\Delta|/(\hbar k_F)$.

This phenomenon is typically known as {\it gapless superconductivity} \cite{deGennes, tinkham2012introduction}.
If we associate  $\tilde{\epsilon}_k$ in Eq. (\ref{Seq:vanishing_epsilon_k}) to the physical energy gap (we remind that $\tilde{\epsilon}_k$ is the {\it ground state energy}), depending on the injected current and $q$, we can have a vanishing energy gap.
In this case, there could be a zero energy gap but the system can be still superconducting since the pairing potential is different from zero \cite{tinkham2012introduction}.
However, it must be stressed that this phenomenon relies on the assumption made about the gap and, in a more general situation, an extreme care must be taken in the determination of the correct and relevant physical gap.

Coming back to the present model, in the chosen gauge, the phase acquired by the pairing potential can be seen as a change in the momentum: $\kvec \rightarrow \kvec + e t ~\Efvec/ \hbar$ and $-\kvec \rightarrow -\kvec + e t ~\Efvec/ \hbar$.
Said otherwise, {\it at a fixed time} $t$, identifying $\qvec = e t ~\Efvec/ \hbar$ we arrive to a situation similar to the one discussed above.

If one wants to compare the full Hamiltonian  (\ref{Seq:H_eff_a}) against the one in Eq. (\ref{Seq:H_spin}), he must take into account the contribution proportional to the identity $\epsilon_k = \sqrt{\xi_k^2 + |\Delta|^2}$.
The full energies are
\begin{eqnarray}
  \epsilon_{k,\pm} &=& \frac{ h_{\kvec_-}-h_{\kvec_+}}{2} \pm \sqrt{  \Big( \frac{h_{\kvec_-}+h_{\kvec_+}}{2}\Big)^2 + |\Delta|^2} = \nonumber \\
  &=& 
  \frac{h_{-\kvec - \frac{e}{c} {\bf A}}-h_{\kvec - \frac{e}{c} {\bf A}}}{2} \pm \sqrt{  \Big( \frac{h_{-\kvec - \frac{e}{c} {\bf A}} + h_{\kvec - \frac{e}{c} {\bf A}}}{2} \Big)^2 + |\Delta|^2}.
  \label{Seq:energy_full}
\end{eqnarray} 
Despite the similarities with Eq. (\ref{Seq:gapless_epsilon_k}), the situation is different as discussed in sec. $5$ of Leggett's book \cite{Leggett_QuantumLiquids} and later in this section.

The energies in Eqs.  (\ref{Seq:gapless_epsilon_k}) and (\ref{Seq:energy_full}) correspond to the ground and excited states.
Since the electric field induces only double quasi-particle excitations and, thus the transition from the ground to the excited state, the relevant excitation gap is the difference between the two energies
\begin{equation}
  \epsilon_{k} = \epsilon_{k,+} - \epsilon_{k,-} = 
   2 \sqrt{  \Big( \frac{h_{-\kvec - \frac{e}{c} {\bf A}} + h_{\kvec - \frac{e}{c} {\bf A}}}{2} \Big)^2 + |\Delta|^2}.
   \label{Seq:non_vanishing_epsilon_k}
\end{equation} 
This never vanishes and the minimum gap is $2 \Delta$.
The time dependence is determined by the kinetic energy and not by the phase of the pairing potential.

As anticipated above, the difference between the present case and the gapless superconductivity relies on the energy gap associated to the physical phenomenon discussed.
Associating the energy gap with the vanishing ground state energy (\ref{Seq:vanishing_epsilon_k}) would correspond to considering the transition from the ground to a single quasi-particle state (which has vanishing energy) \cite{Leggett_QuantumLiquids}.
This is the situation in which a thermal-like excitation breaks a Cooper pair (in Leggett's terminology they are called pair-breaking transitions \cite{Leggett_QuantumLiquids}).
On the contrary, the non-vanishing gap (\ref{Seq:non_vanishing_epsilon_k}) corresponds to the ground-to-excited state transition that are the only ones that can occur in presence of an electric field.

In the physical situation treated in the paper, the thermal pair-breaking effects can be neglected in the time-scale over which the excitation takes place. In fact, since the temperature is much smaller than the critical temperature, it is natural to assume that the thermal excitations are suppressed.
We conclude that, since the electric field creates two quasi-particles, the relevant energy gap is the one in Eq. (\ref{Seq:non_vanishing_epsilon_k}), and consequently it  can not vanish. Thus, there is no a gapless superconductivity phenomenon.

As a final remark, we would like to point out that the time-dependent phase acquired by the pairing potential depends on the gauge choice and, therefore, it cannot be the origin of physical effects.
On the contrary, in our numerical simulation we solve the Schroedinger equation and calculate the physical observables (such as the ground and excited state populations) which are always gauge independent.

\end{document}